\documentclass[%
 aip,
 amsmath,amssymb,
 reprint,%
]{revtex4-1}

\usepackage{graphicx}
\usepackage{dcolumn}
\usepackage{bm}

\usepackage[utf8]{inputenc}
\usepackage[T1]{fontenc}
\usepackage{mathptmx}
\usepackage{etoolbox}
\usepackage{siunitx}
\makeatletter
\def\@email#1#2{%
 \endgroup
 \patchcmd{\titleblock@produce}
  {\frontmatter@RRAPformat}
  {\frontmatter@RRAPformat{\produce@RRAP{*#1\href{mailto:#2}{#2}}}\frontmatter@RRAPformat}
  {}{}
}%
\makeatother
\begin{document}

\preprint{AIP/123-QED}

\title[Polarization-multiplexed erbium dual-comb fiber laser]{Detection of carbon monoxide using a polarization-multiplexed erbium dual-comb fiber laser}
\author{P. E. Collin Aldia}%
 \email{collinaldia@univie.ac.at, oliver.heckl@univie.ac.at}
\affiliation{Optical Metrology Group, Faculty of Physics, University of Vienna, Boltzmanngasse 5, 1090 Vienna, Austria}%
\affiliation{Vienna Doctoral School in Physics, University of Vienna, Boltzmanngasse 5, 1090 Vienna, Austria}%
\author{Jiayang Chen}
\affiliation{Optical Metrology Group, Faculty of Physics, University of Vienna, Boltzmanngasse 5, 1090 Vienna, Austria}
\affiliation{State Key Laboratory of Precision Measurement Technology and Instruments, Department of Precision Instrument, Tsinghua University, Beijing 100084, China}
\author{Jonas K. C. Ballentin}
\affiliation{Optical Metrology Group, Faculty of Physics, University of Vienna, Boltzmanngasse 5, 1090 Vienna, Austria}

\author{Lukas W. Perner}
 \altaffiliation[Currently at the ]{Laboratoire Temps-Fréquence, Institut de Physique, Université de Neuchâtel}%
\affiliation{%
Christian Doppler Laboratory for Mid-IR Spectroscopy and Semiconductor Optics, Faculty Center for Nano Structure Research, Faculty of Physics, University of Vienna, Boltzmanngasse 5, 1090 Vienna, Austria
}%
\author{O. H. Heckl}%
\affiliation{Optical Metrology Group, Faculty of Physics, University of Vienna, Boltzmanngasse 5, 1090 Vienna, Austria}%




\date{\today}

\begin{abstract}
We present a compact, reliable, and robust free-running all-polarization-maintaining erbium (Er) single-cavity dual-comb laser generated via polarization multiplexing with gain sharing. Polarization multiplexing exploits the fast and slow axes of the fiber, while modelocking is achieved through a nonlinear amplifying loop mirror scheme using readily available components. The laser operates at a repetition rate of around \SI{74.74}{MHz} with a tuning capability in the difference in repetition rates from \SI{500}{Hz} to \SI{200}{kHz}. This tunability makes the system more flexible for dual-comb spectroscopy experiments. Consequently, using this laser, we demonstrated a proof-of-principle dual-comb spectroscopy of carbon monoxide (CO), operating without any active stabilization. 
\end{abstract}

\maketitle

%

\section{\label{sec:level1}Introduction}

Optical Frequency Combs (OFCs) have become ubiquitous tools in various fields, among them molecular fingerprinting,\cite{diddams2007molecular} trace gas sensing,\cite{alden2017methane} precision ranging,\cite{minoshima2000high} atomic clocks,\cite{rosenband2008frequency} and high-harmonic generation.\cite{baltuvska2003attosecond} In particular, owing to their high coherence, high-frequency resolution, and broadband spectra, OFCs are considered to be an ideal tool/light source for molecular spectroscopy. \cite{picque2019frequency,coddington2016dual,shumakova2024short} This technique, called direct frequency comb spectroscopy, has enabled groundbreaking results in fields as diverse as time-resolved spectroscopy, \cite{ideguchi2013coherent,lomsadze2018tri} human breath analysis, \cite{thorpe2006broadband} and real-time sensing. \cite{rieker2014frequency} Conventionally, Fourier transform spectrometers (FTS),\cite{bates1976fourier,maslowski2016surpassing} grating-based spectrometers, and virtually imaged phase arrays (VIPAs) \cite{shirasaki1996large} are used as spectrometers. However, these systems require bulky setups, long scanning ranges, and rigorous alignment. 

In contrast, so-called dual-comb (DC) systems eliminate all the aforementioned problems of conventional spectrometers. DC spectroscopy combines the advantages of the above-mentioned single-comb techniques by leveraging the properties of optical frequency combs. In essence, a DC system consists of two mutually coherent modelocked lasers with spectral overlap with slightly different pulse repetition rates. In the frequency domain, this results in two frequency combs with slightly different comb line spacing. The optical beating between the comb lines of these two OFCs generates a down-converted radio frequency (RF) signal. Conveniently, only a fast photodiode is necessary to detect this resulting RF beat signal, making the above-mentioned complex spectrometer assemblies unnecessary. This RF interferogram contains relevant spectral information in the optical domain. Thus, high-resolution spectra can be obtained at an acquisition rate considerably higher than that of single-comb spectroscopy (mostly limited by the difference in repetition rates of the two combs) without the need for any sophisticated measurement tools. This technique was first proposed by Schiller \textit{et al.},\cite{schiller2002spectrometry} and later demonstrated by Coddington \textit{et al.} \cite{coddington2008coherent,coddington2016dual} Thereafter, DC systems have found a wide range of applications, notably in ranging, \cite{coddington2009rapid,fellinger2021simple} absorption spectroscopy \cite{tian2022dual} and CARS. \cite{ideguchi2013coherent,qin2021all} Since the two combs have to be mutually coherent (i.e., retain a stable relation between Comb1 and Comb2), they need to be stabilized to each other. This effectively shifts the complexity of the detection apparatus to the laser source, again leading to complicated setups. Several ways to overcome this challenge were proposed, among them, phase locking the two combs to an external cavity diode laser \cite{coddington2008coherent,truong2016accurate} or using adaptive sampling techniques. \cite{giaccari2008active,ideguchi2014adaptive} An elegant solution is so-called single-cavity dual-comb (SCDC) lasers, where both combs are generated in a single laser cavity. This guarantees good mutual coherence due to common-mode noise cancellation. Consequently,  applications without comb-mode resolution do not require active stabilization. Since the advent of SCDC, several schemes have been contrived, among them bi-directional splitting, \cite{ideguchi2016kerr,galtier2022high} branched optical paths in a birefringent crystal, \cite{gu2023polarization,willenberg2020femtosecond} spatial separation,\cite{pupeikis2022spatially} polarization multiplexing, \cite{zhao2018polarization,sterczewski2019computational} or spectral subdivision, commonly called dual-color systems.\cite{fellinger2019tunable,fellinger2019tunable1,zhao2011switchable,liao2018dual}

\begin{figure*}
\centering\includegraphics[width=0.95\linewidth]{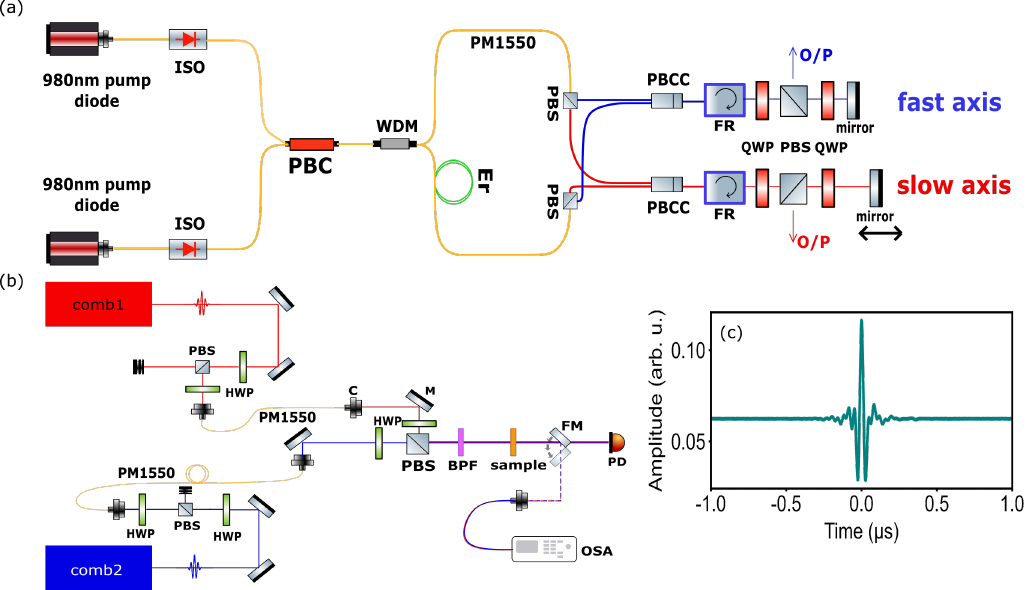}
\caption{(a) The Er dual-comb laser setup. ISO: isolator; PBC: polarizing beam combiner; WDM: wavelength division multiplexer; Er: Er-doped gain fiber; PBS: polarizing beam splitter; PBCC: polarizing beam combiner collimator; FR: Faraday rotator; QWP: quarter-wave plate. (b) Experimental setup to obtain the optical beating between the two combs. HWP: half-wave plate; BPF: bandpass filter; C: collimator; M: mirror; FM: flip mirror; PD: photodiode; OSA: Optical Spectrum Analyzer. (c) A single-shot interferogram recorded using the setup in (b).}
\label{figure 1}
\end{figure*}

All of the SCDC systems mentioned above are based on modelocked lasers. Particularly, OFCs based on polarization maintaining (PM) fibers modelocked via nonlinear amplifying loop mirror (NALM) are known to be more robust and reliable, also offering extremely low free-running noise properties when operated in certain regimes. \cite{mayer2020flexible,kuse2016all,hansel2018all} Fellinger \textit{et al.} \cite{fellinger2019tunable} and Nakajima \textit{et al.} \cite{nakajima2021mechanical} demonstrated DC systems based on PM fibers modelocked using a NALM. The former is a dual-color system based on spectral subdivision while the latter exploits mechanical sharing of components. In the former setup, additional spectral broadening is required to overlap the dual-color laser and obtain an RF beat signal. Other approaches include polarization multiplexing to develop a DC system using a NALM modelocking scheme. \cite{nakajima2019all,rao2022polarization} More recently, Iwakuni \textit{et al.} \cite{iwakuni2023noise} studied the noise properties of an SCDC modelocked using a NALM based on polarization multiplexing with gain sharing. They reported a low phase noise for their system, making it a suitable candidate for DC spectroscopy. This type of DC system with polarization splitting has inherent spectral overlap, making spectral broadening unnecessary.

Here, we present a PM fiber-based polarization-multiplexed SCDC laser, modelocked via NALM with gain sharing. In free-running dual-comb operation (i.e., without any active stabilization), the difference in repetition rate between the two combs $\Delta$\textit{f}$_\mathrm{rep}$ was tuned over almost three orders of magnitude, from \SI{500}{Hz} to \SI{200}{\kilo\hertz} while maintaining stable dual-comb operation. This tuning possibility gives us a convenient handle on the acquisition rate, which directly depends on $\Delta$\textit{f}$_\mathrm{rep}$ and the non-aliasing bandwidth $\Delta\nu$. The latter is the maximum allowed overlap between the two combs before spectral aliasing occurs. It is given by \cite{ideguchi2017dual}

\begin{equation}
\Delta\nu \leq \frac{\textit{f}_\mathrm{rep,1}^{~2}}{2\Delta\textit{f}_\mathrm{rep} }.
\label{eq 1}
\end{equation}

 With the possibility to tune $\Delta$\textit{f}$_\mathrm{rep}$ between \SI{500}{Hz} to \SI{200}{\kilo\hertz}, we were able to achieve a non-aliasing bandwidth of \SIrange{0.014}{5.59}{THz} (\SIrange{0.12}{45.93}{nm} at \SI{1570}{nm}). While modelocking could be maintained over the whole tuning range, we chose a $\Delta$\textit{f}$_\mathrm{rep}$ of \SI{1.04}{kHz} for our proof-of-principle DC spectroscopy demonstration, resulting in a $\Delta\nu$ of \SI{2.69}{THz} at \SI{1570}{nm}. This value of $\Delta$\textit{f}$_\mathrm{rep}$ is an optimal compromise between fast acquisition rate and large non-aliasing bandwidth for our target application. Our system is based on the aforementioned NALM technique and polarization multiplexing; it has inherent spectral overlap and is compact and robust without active stabilization, making it an ideal basis for field-deployable spectroscopy. 

\begin{figure*}
\centering\includegraphics[width=0.85\linewidth]{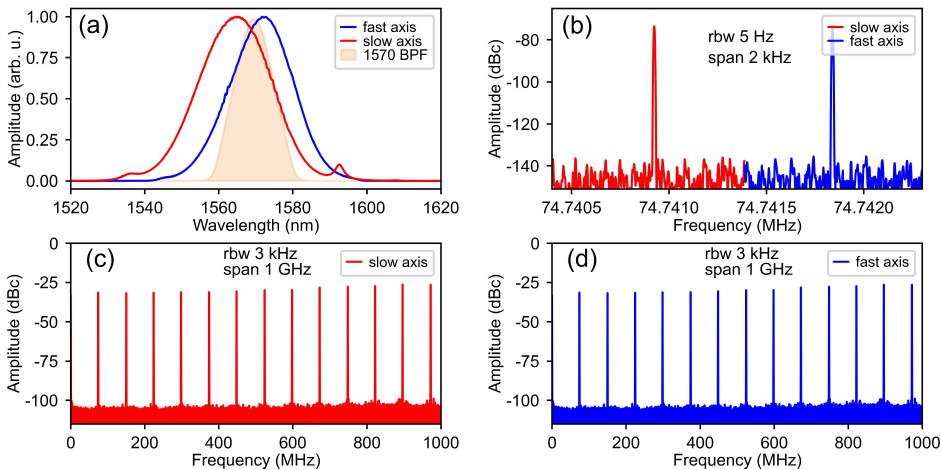}
\caption{(a) Output spectra of the two combs measured using an optical spectrum analyzer (Yokogawa AQ6374). The shaded region shows the transmission curve of the bandpass filter. (b) The radio frequency spectrum of the laser was measured using a fast photodiode. This plot shows a single measurement and the two colors are just a guide for visual representation. (c),(d) \SI{1}{GHz} span and a resolution bandwidth (rbw) of \SI{3}{kHz} of the individual combs showing their repetition rate and higher harmonics for fast and slow axes, respectively. All the RF spectra were recorded using an RF spectrum analyzer (Keysight PXA N9030B).}
\label{figure 2}
\end{figure*}
 
\section{\label{sec:level2}Experimental setup}

As shown in Fig. \ref{figure 1}(a), the laser setup consists of an all-PM-fiber section and two free-space sections. The fiber section comprises a PM wavelength division multiplexer (WDM), an Er-doped gain fiber (Liekki Er80-4/125-HD-PM) with a length of \SI{78}{cm}, two polarizing beam splitters (PBS), and two birefringent polarizing beam combiner collimators (PBCC). Since Er-doped gain fibers have less gain compared to ytterbium-doped gain fibers, two \SI{980}{nm} pump diodes (3SP 3CN01800JL) are used for initiating DC operation. The pump light is combined at a polarizing beam combiner (PBC) which then travels through the common port of the WDM to the gain fiber. The light exiting the PBCC travels through various polarization optics in the free-space section. 

The modelocking scheme used here is similar to Ref. \onlinecite{mayer2020flexible}. The polarization of light in the fiber ring is split at the fiber-PBS into the fast and slow axes of the PM fiber. It is at this fiber component where polarization multiplexing of DC operation begins. After the splitting, the light exits the respective PBCCs into the free-space arms. The non-reciprocal phase bias introduced by the Faraday rotator and the quarter-waveplate, along with the free-space PBS and the other quarter-waveplate initiate modelocking the laser.

The two combs share the gain medium and also \SI{116}{cm} of single-mode PM1550 fiber. The unshared fiber paths have a length of \SI{48.4}{cm} and \SI{48.75}{cm} for slow and fast axes, respectively. The linear free-space arms have virtually identical lengths around \SI{13.5}{cm}. Additionally, one of the end mirrors is mounted on a translational stage to fine-tune the difference in repetition rates $\Delta\textit{f}_\mathrm{rep}$. To start the DC operation, we first tune the waveplate positions for each arm individually (i.e., with the other arm blocked) to obtain self-starting states. Once these states have been found, we increase the pump currents to the sum of the pump currents required for modelocking each comb individually. At this value of pump current, we adjust the linear losses in the two arms by rotating the quarter-waveplate between the PBS and the end mirror. The dual-comb operation can be initiated by ensuring that linear losses in the two arms are comparable. Afterward, we gradually decrease the pump current until the single-pulse operation is achieved for both combs. At this point, only one of the pump diodes is required to maintain the DC state. We ensure single-mode operation by verifying the absence of multiple interferograms on an oscilloscope trace spanning $\Delta\textit{t} = 1/\Delta\textit{f}_\mathrm{rep}$. Once this is achieved, the system stays in dual-comb operation for several months.

\begin{figure*}
\centering\includegraphics[width=0.95\linewidth]{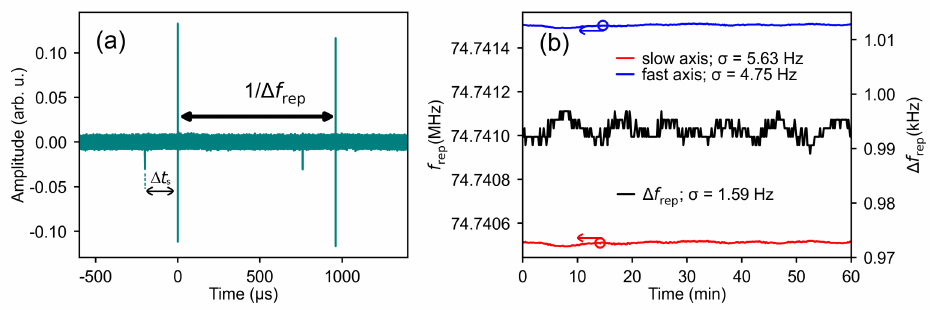}
\caption{(a) Time trace recorded on an oscilloscope (LeCroy WavePro 760Zi) where the center-bursts are separated in time $\Delta t=1/\Delta\textit{f}_\mathrm{rep}$. The spurious signals resulting from the intracavity pulse collisions are visible as small side-peaks (discussed in section \ref{sec:level3a}). (b) The drift of the repetition rates and their difference $\Delta\textit{f}_\mathrm{rep}$.}
\label{figure 3}
\end{figure*}

\section{\label{sec:level3}Dual-comb characterization}

Our polarization-multiplexed Er single-cavity dual-comb laser has a nominal repetition rate around \SI{74.74}{MHz} with a difference in repetition rates $\Delta\textit{f}_\mathrm{rep}$ tunable from \SI{500}{Hz} to \SI{200}{kHz}. As shown in Fig. \ref{figure 2}(a), the output spectra from the two combs are extremely similar as expected for a DC system based on polarization multiplexing. The spectra are centered around \SI{1571.8}{nm} and \SI{1564.6}{nm} with a full width at half maximum (FWHM) of \SI{20.09}{nm} and \SI{23.55}{nm} for fast and slow axes, respectively. We attribute the discrepancy in center wavelength to a difference in linear losses among the two arms.\cite{Paschottaerbium_doped_laser_gain_media} The net dispersion of the cavity is calculated to be $\sim$ -\SI{0.015}{ps}$^{2}$ at their center wavelengths. 
The radio frequency spectrum of the respective pulse trains is measured using a fast photodiode (Thorlabs DET08CL/M) as shown in Fig. \ref{figure 2}(c)-(d). The higher harmonics can also be seen in these plots and the absence of side peaks around the main peaks confirms the clean single-pulse operation. In this state, the output power from the laser was directly measured to be \SI{14}{mW} and \SI{11}{mW} for fast and slow axes, respectively. To determine the repetition rates of the laser, $\textit{f}_\mathrm{rep,1}$ and $\textit{f}_\mathrm{rep,2}$, and the difference in repetition rates $\Delta\textit{f}_\mathrm{rep}$, the output from the two arms are sent to a Thorlabs PDA05CF2 photodiode (Fig. \ref{figure 2}(b)).

To record the beat signal (the interferogram as shown in Fig. \ref{figure 1}(c)), the two combs are spatially overlapped and the signal is measured using a simple photodiode (Thorlabs PDA05CF2) (Fig. \ref{figure 1}(b)). A \SI{48}{MHz} low-pass filter is used to remove the repetition rates of the laser $\textit{f}_\mathrm{rep,1}$ and $\textit{f}_\mathrm{rep,2}$ before the signal is recorded on an oscilloscope (LeCroy WavePro 760Zi). Calculating the Fast-Fourier-Transform (FFT) of this interferogram yields the down-converted RF spectrum. The center frequency of this down-converted spectrum is associated with the difference in the carrier-envelope offset (CEO) frequency. However, this parameter is not easy to access, and it is, therefore, difficult to shift the down-converted spectrum in the RF domain deliberately. This effectively limits the extent to which the full non-aliasing bandwidth can be used for spectroscopy experiments. Nevertheless, the non-aliasing bandwidth was large enough for this work with our specified laser parameters. As this paper is focused on the capabilities of the system in free-running operation, controlling the CEO frequency (as well as dispersion tuning) is left to future investigation.

\subsection{\label{sec:level3a}Intra-cavity cross-talk and resolution}

When two pulse trains travel within the same cavity or share some parts of the cavity, they interact with each other.\cite{wei2018ultrafast} This causes the formation of spurious signals in the time traces. As seen in the work of Fellinger \textit{et al.}, \cite{fellinger2019tunable} a simple delay line in one of the arms conveniently shifts these spurious signals with respect to the interferogram before the spatial overlap of the two combs. Thus, by creating an appropriate extra-cavity path length between the two arms, we can achieve an undisturbed time window around the center burst. This is relevant for dual-comb spectroscopy since the resolution in the radio frequency domain directly corresponds to the inverse of the measured time window. For \SI{1}{m} of fiber inserted into one of the arms delays the spurious signal by $\sim$\SI{200}{\micro\second} (Fig. \ref{figure 3}(a)). This would correspond to an undisturbed time window of \SI{400}{\micro\second} around the center burst. For a nominal repetition rate around \SI{74.74}{MHz} and $\Delta\textit{f}_\mathrm{rep}$ of \SI{1.04}{kHz}, this time window would accord to an optical resolution $\sim$\SI{180}{MHz} (a resolution of \SI{0.01}{nm} at \SI{1570}{nm}).

\subsection{\label{sec:level3b}Non-aliasing bandwidth}

As mentioned in section \ref{sec:level1}, the non-aliasing bandwidth (Eq. \ref{eq 1}) is an important parameter, as it gives the maximal allowed spectral overlap. Therefore, once the two arms are spatially overlapped, we insert a bandpass filter with an FWHM of \SI{12}{nm} to avoid aliasing. For the above laser parameters, the non-aliasing bandwidth is calculated to be \SI{22.13}{nm} at \SI{1570}{nm}. Since the non-aliasing bandwidth depends on the repetition rate of the laser and the difference in repetition rates, we investigated the tuning possibility of $\Delta\textit{f}_\mathrm{rep}$. We were able to tune the $\Delta\textit{f}_\mathrm{rep}$ from \SI{500}{Hz} to \SI{200}{kHz} above which we observe a reduction of $\Delta\nu$ and mutual coherence. Conversely, reducing the lower bound will compromise the maintenance of dual-comb operation due to injection locking. Even though a $\Delta\textit{f}_\mathrm{rep}$ of \SI{500}{Hz} would give a large value of $\Delta\nu$, the long-term stability of the laser was compromised. However, $\Delta\textit{f}_\mathrm{rep}$ of around \SI{1}{kHz} ensured better performance in terms of stability over several months and also a reasonable value of $\Delta\nu$.

\begin{figure*}
\centering\includegraphics[width=0.95\linewidth]{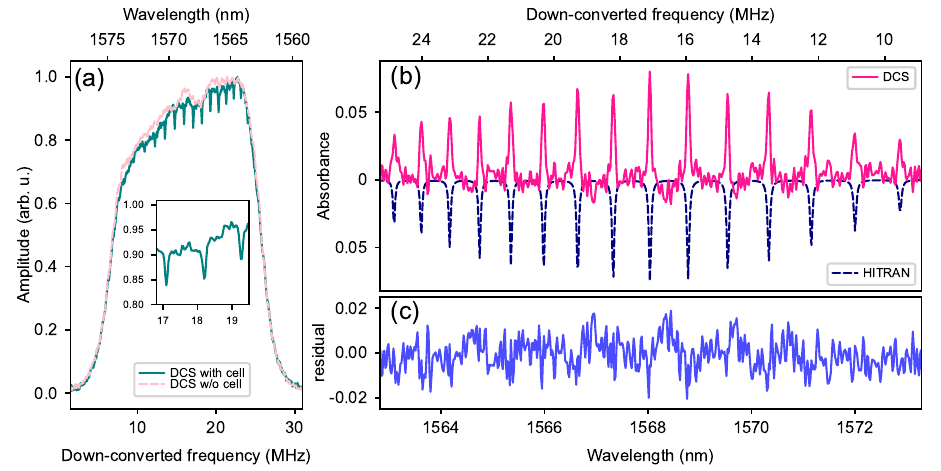}
\caption{Dual-comb spectroscopy of carbon monoxide. The difference in repetition rates of the laser is set to \SI{1.04}{kHz}. (a) The down-converted RF spectrum was obtained after performing FFT of an average of 250 interferograms with CO cell (teal) and without cell (dashed pink). (b) The deep pink line shows the absorption spectrum from DC detection while the blue-dashed lines depict the HITRAN curve. The HITRAN data is calculated for a path length of \SI{30}{cm}, \SI{790}{Torr} and \SI{296}{K}. (c) The residuals (HITRAN - DC) from (b) after fitting.}
\label{figure 4}
\end{figure*}

\subsection{\label{sec:level3c}Frequency stability}

The frequency stability of this dual comb system was measured in free-running operation without any active stabilization. To carry out this measurement, we observed the drifts of the two individual repetition rates for one hour (Fig. \ref{figure 3}(b)). However, the drift in $\Delta\textit{f}_\mathrm{rep}$ was calculated by the difference of $\textit{f}_\mathrm{rep,1}$ and $\textit{f}_\mathrm{rep,2}$, and not measured directly (Fig. \ref{figure 3}(b); black line). The value of $\Delta\textit{f}_\mathrm{rep}$ for this measurement was set close to \SI{1}{kHz}. The drift in the magnitude of the individual repetition rates was estimated by calculating the standard deviation $\sigma$ which was computed to be \SI{5.63}{Hz} and \SI{4.75}{Hz} for slow and fast axes, respectively. However, the measured drift in $\Delta\textit{f}_\mathrm{rep}$ shows a reduction by a factor of three with $\sigma$ = \SI{1.59}{Hz}. This value can be further reduced by improving the mechanical stability of the laser or proper boxing of the system. Nevertheless, the value of the drift was already good enough to carry out proof-of-principle measurements with this dual comb system.

\section{\label{sec:level4}Dual-comb spectroscopy of Carbon Monoxide ($^{12}$C$^{16}$O)}
In this section, we demonstrate a proof-of-principle dual-comb spectroscopy experiment of a carbon monoxide (CO) reference cell at ambient temperature. The absorption features of the gas are measured at a pressure of \SI{790}{Torr} with an effective path length of \SI{30}{cm}. For these measurements, a time window of $\SI{20}{\si{\micro}s}$ was set, and 250 interferograms were recorded in sequence using an oscilloscope (LeCroy WavePro 760Zi). This time window corresponds to an optical frequency resolution of \SI{3.6}{GHz} for $\Delta\textit{f}_\mathrm{rep}$ = \SI{1.04}{kHz} and $\textit{f}_\mathrm{rep}$ = \SI{74.74}{MHz}. These 250 interferograms were averaged in the time domain after applying a phase correction algorithm from Ref. \onlinecite{phillips2023coherently}. Afterward, an FFT was performed to obtain the down-converted spectrum (Fig. \ref{figure 4}(a)). The same procedure is carried out without the cell to obtain a background spectrum also shown in Fig. \ref{figure 4}(a). Subtracting the resulting spectrum with cell from the background spectrum, we obtain the absorption spectrum of CO. The wavelength calibration for this down-converted RF spectrum is carried out using a calibrated grating-based optical spectrum analyzer (Yokogawa AQ6374).  Afterward, the absorption spectrum is compared with the reference data from HITRAN with the same temperature, pressure, and resolution as that of DC. \cite{borkov2020fourier} Figure \ref{figure 4}(b) shows the resulting absorption spectrum (deep pink) of CO after background subtraction and baseline correction of the DC system. The absorption lines detected by the DC system exhibit good agreement with those derived from HITRAN (blue-dashed lines) as depicted in Fig. \ref{figure 4}(b). The residuals of the experiment are shown in Fig. \ref{figure 4}(c). It is calculated by the difference between HITRAN and DC spectroscopic data. It is to be noted that the value of the residuals is within the noise floor of the measurement system. This shows the spectroscopic capability of this dual-comb system even in free-running operation.

\section{\label{sec:level5}Conclusion and outlook}

We have demonstrated a method to build an all-PM Er single-cavity dual-comb laser based on polarization multiplexing. This approach leverages the fast and slow axes of the PM fiber for subdivision while using a NALM to achieve stable modelocking. As demonstrated by our subsequent proof-of-principle measurements of CO absorption features, these factors result in a reliable and robust source for DC spectroscopy. Since both combs share the same gain medium and pump diode, most of the system noise is common mode, thereby enhancing mutual coherence. Compared to previous work on gain sharing, \cite{iwakuni2023noise} we increased the non-aliasing bandwidth by a factor of $\sim$150 and the average power by a factor of 6. Moreover, having one of the end mirrors on a translational stage makes it easier to tune the difference in the repetition rates of the laser. By means of this translatable mirror, we were able to tune the $\Delta\textit{f}_\mathrm{rep}$ from \SI{500}{Hz} to \SI{200}{kHz}. This plays a crucial role in the trade-off between the acquisition time and the non-aliasing bandwidth which becomes relevant for DC spectroscopy. Notably, the system stayed in stable DC operation for several months.

The DC spectroscopic capability of this system was demonstrated as a proof-of-principle experiment by measuring the transmission curve of a carbon monoxide reference cell. After data processing, including background and baseline correction, as well as a comparison to HITRAN reference data, we were able to resolve absorption peaks with an optical resolution of \SI{3.6}{GHz} in a completely free-running operation. Although the noise floor of the DC system needs improvement, the residuals remain within the noise floor. Enhancement of the SNR will be addressed in future works, for instance, by using real-time data processing. \cite{ideguchi2014adaptive} This proof-of-principle detection of CO and the dual-comb scheme make this system a promising candidate for the development of compact, robust, and high-resolution spectrometers tailored for field-deployable spectroscopy. 

\begin{acknowledgments}
We thank Maximillian Prinz, Zbigniew Łaszczych, and Monika Bahl for valuable discussions. This research was funded by the Austrian Science Fund (FWF) [10.55776/P33680]. The financial support by the Austrian Federal Ministry for Digital and Economic Affairs, the National Foundation for Research, Technology and Development, and the Christian Doppler Research Association is gratefully acknowledged.

Jiayang Chen is a visiting student who would like to acknowledge the support from the China Scholarship Council.

Certain instruments are identified in this paper in order to specify the experimental procedure adequately. Such identification is not intended to imply recommendation or endorsement by all the authors, nor is it intended to imply that the instruments identified are necessarily the best available for the purpose.
\end{acknowledgments}

\section*{Author Declarations}
 \subsection*{Conflict of Interest}
 The authors have no conflicts to disclose.
 \subsection*{Data Availability Statement}
The data that support the findings of this study are available from the corresponding authors upon reasonable request.

\section*{\label{sec:Inst}References}

\bibliography{references}

\end{document}